 \PassOptionsToPackage{colorlinks=true}{hyperref}

\documentclass[onecolumn,sn-mathphys-num]{sn-jnl}

\usepackage{bm}
\usepackage{breqn}
\usepackage{amssymb}
\usepackage{latexsym}
\usepackage{amsfonts}
\usepackage{mathrsfs}
\usepackage{stackrel}
\usepackage{setspace}
\usepackage{stackengine}
\usepackage{graphicx}
\usepackage{tabularx}
\usepackage{multirow}
\usepackage{epsfig}
\usepackage{epstopdf}
\usepackage{subcaption}
\usepackage{booktabs} 

\numberwithin{equation}{section}

\newcommand{\norm}[1]{\left\lVert#1\right\rVert}

\newtheorem{remark}{Remark}[section]

\begin{document}


\title[Article Title]{A high order accurate and provably stable fully discrete continuous Galerkin framework on summation-by-parts form for advection-diffusion equations}
\author[1]{\fnm{Mrityunjoy} \sur{Mandal}}\email{mrityunjoy.mandal@uct.ac.za}
\author*[3,2]{\fnm{Jan} \sur{Nordstr\"{o}m}}\email{jan.nordstrom@liu.se}
\author[1]{\fnm{Arnaud} G\sur{Malan}}\email{arnaud.malan@uct.ac.za}
\affil[1]{\orgdiv{InCFD Research Group, Department of Mechanical Engineering}, \orgname{University of Cape Town},\orgaddress{ \city{Cape Town},\country{South Africa}}}

\affil*[2]{\orgdiv{Department of Mathematics}, \orgname{Link\"{o}ping University},\orgaddress{\city{Link\"{o}ping},  \country{Sweden}}}
\affil[3]{\orgdiv{Department of Mathematics and Applied Mathematics}, \orgname{University of Johanesburg},\orgaddress{ \city{Johanesburg},\country{South Africa}}}


\abstract{
We present a high-order accurate fully discrete numerical scheme for solving Initial Boundary Value Problems (IBVPs) within the Continuous Galerkin (CG)-based Finite Element framework. Both the spatial and time approximation in Summation-By-Parts (SBP) form are considered here. The initial and boundary conditions are imposed weakly using the Simultaneous Approximation Term (SAT) technique. The resulting SBP-SAT formulation yields an energy estimate in terms of the initial and external boundary data, leading to an energy-stable discretization in both space and time. The proposed method is evaluated numerically using the Method of Manufactured Solutions (MMS). The scheme achieves super-convergence in both spatial and temporal direction with accuracy $\mathcal{O}(p+2)$ for $p\geq 2$, where $p$ refers to the degree of the Lagrange basis. In an application case, we show that the fully discrete formulation efficiently captures space-time variations even on coarse meshes, demonstrating the method’s computational effectiveness. 
}


\maketitle

\section{Introduction}   
In this work, we present a highly accurate, fully discrete, and energy-stable numerical scheme based on the SBP–SAT technique \cite{svard2014review,fernandez2014review}. The transient advection–diffusion equation is considered for various flow regimes, ranging from diffusion-dominated to advection-dominated cases. The SBP–SAT framework in the CG formulation was previously developed \cite{hanif2025efficiency} for spatial discretizations, where a convergence rate of order $\mathcal{O}(p+2)$ was observed when an explicit Runge–Kutta time integration scheme was employed with a sufficiently small time step. In the present work, we extend this framework to the temporal domain \cite{nordstrom2013summation,lundquist2014sbp}. For time discretization, both global and one-step multi-stage approaches are employed. The SAT technique is used to weakly impose both the initial and boundary conditions. The resulting formulation is provably stable and exhibits superconvergent behavior in both the temporal and spatial directions. To the best of our knowledge, such a highly accurate, fully discrete, and provably stable time integration scheme is not previously found within the CG framework.

For time integration of stiff IBVPs, implicit schemes are often preferred, as they relax the stability requirement on the time step size. Common examples of such methods include implicit Runge–Kutta schemes \cite{kennedy2003additive,carpenter2005fourth}, Backward Differentiation Formulas (BDF) \cite{cash1983integration,cash2000modified}, and linear multistep methods \cite{hundsdorfer2006monotonicity,hundsdorfer2012stepsize}. These approaches can be regarded as local methods since each time interval depends only on one or a few previous steps. In contrast, global methods consider the entire time domain, ranging from the initial time to the final time $T$. Previously developed global methods, such as spectral and collocation time approximations \cite{costabile2001method,guo2009legendre,wang2012legendre}, are often considered impractical due to their high computational costs. Nevertheless, global methods exhibit unconditional stability and can achieve very high orders of accuracy. Moreover, they can be used to exactly mimic continuous energy estimates, a property that is rarely achieved by local methods. SBP in time \cite{nordstrom2013summation} was introduced as a global method and was later \cite{lundquist2014sbp} reformulated as a multistage method, retaining its global stability properties. 


 The paper is organized as follows. The continuous problem and the associated energy analysis are described in Section~\ref{sec:MathematicalFormulations}. The appropriate choice of boundary conditions and their weak imposition using SATs are also presented. The semi-discrete approximation of the continuous problem and its associated energy estimate are discussed in Section~\ref{Sec:Semidiscrete_CGFEM}. Here, we also present the discrete energy approximation and prove stability in a manner similar to that of the continuous setting. The fully discrete scheme is discussed in detail, along with the energy analysis in Section~\ref{Sec:FullyDiscrete_approximation}. The one-step multistage technique is discussed in Section~\ref{Sec:MultiStage_TimeIntegration} and is also shown to be energy stable. Section~\ref{Sec:Numerical_results_advectDiff_results} presents the numerical solutions for three different problems. First, we use the MMS to verify the order of temporal convergence of the formulation. Secondly, we evaluate the order of spatial convergence using another MMS, and lastly, we evaluate the performance of the combined effort by solving a wave propagation problem. Conclusions are presented in Section~\ref{sec:Concluding_remarks}.     
 
\section{The continuous problem }
\label{sec:MathematicalFormulations}
Let us consider a one-dimensional domain $x\in\Omega=[x_{0},x_{1}]=[0,1]$ with the IBVP:
\begin{equation}
\label{Eq:AdvectDiff_GE}
\begin{split}
\partial_{t}u + a\partial_{x}u - \epsilon\partial_{xx}u& = \ 0, \quad\quad 0\leq x\leq 1,\quad  t\geq 0,\\
au-\epsilon\partial_{x}u &= g_{0}(t), \quad\ \ x=0,\quad\quad t\geq 0,\\
\epsilon \partial_{x}u &= g_{1}(t), \quad\ \ \ x=1,\quad \quad t\geq 0,\\
u&=f(x),  \quad 0\leq x\leq 1,\ \quad t=0.
\end{split}
\end{equation}
In (\ref{Eq:AdvectDiff_GE}), $u=u(x,t)$ defines the solution field, and the constants $a>0,\epsilon>0$ denote the advection and diffusion coefficients, respectively. The partial differential operators $\partial_{t},\partial_{x}$ and $\partial_{xx}$ define the temporal and spatial differentiations, respectively. The boundary operators are $(a-\epsilon\partial_{x})$ and $\epsilon\partial_{x}$, the boundary data are $g_{0}$ and $g_{1}$, while the initial data is represented by $f$. 

\subsection{Continuous energy analysis}
\label{Sec:ContEnergyAnalysis}
By multiplying (\ref{Eq:AdvectDiff_GE}) with the solution and performing integration-by-parts in space, followed by the imposition of boundary conditions, we derive the following energy rate: 
\begin{equation}
\label{Eq:AdvectDiff_ContEnergyRate}
\frac{d}{dt}\norm{u}^{2} + 2\epsilon \norm{\partial_{x}u}^{2}
= \dfrac{1}{a}\left.\left[ g_{0}^{2}-\left(au-g_{0}\right)^{2}\right]\right|_{x=0} + \dfrac{1}{a}\left.\left[ g_{1}^{2}-\left(au-g_{1}\right)^{2}\right]\right|_{x=1}\leq \dfrac{1}{a}\left(g_{0}^{2} + g_{1}^{2} \right),
\end{equation}
where $\norm{u}^{2}=\int_{\Omega}u^{2}d\Omega$ and $\norm{\partial_{x}u}^{2}=\int_{\Omega}(\partial_{x}u)^{2}d\Omega$. Time integration of (\ref{Eq:AdvectDiff_ContEnergyRate}) generates the energy estimate:
\begin{equation}
\label{Eq:AdvectDiff_ContEnergyEstStrongBC}
\norm{u(:,T)}^{2}+2\epsilon\int_{0}^{T}\norm{\partial_{x}u}^{2}dt=\norm{f}^{2}+\dfrac{1}{a}\int_{0}^{T}\left[(g_{0}^{2}+g_{1}^{2})-\left[\left(au-g_{0} \right)^{2} + \left(au-g_{1} \right)^{2}\right]\right]dt,
\end{equation}
where the norm of the solution at final time and the time integration of the gradient of the solution are bounded by initial and boundary data. However, the strong imposition of boundary conditions may result in stability issues in the discrete setting \cite{strand1994summation}. In remedy, we will implement the boundary conditions in weak sense using penalty terms.

\subsection{Weak imposition of boundary conditions for the continuous problem}
\label{Sec:WeakImpositionBC}
Following \cite{nordstrom2017roadmap}, we impose the boundary conditions  in (\ref{Eq:AdvectDiff_GE}) weakly by adding  the SAT-like terms as:
\begin{equation}
\label{Eq:WeakImpositionBC_AdvectDiff}
\begin{split}
\partial_{t}u + a\partial_{x}u-\epsilon\partial_{xx}^{2}&=\textrm{SAT} \; \textrm{where} \; \textrm{SAT} = L_{0}\left(\sigma_{0}\left[\left(au-\epsilon\partial_{x}u \right)-g_{0} \right]\right) + L_{1}\left(\sigma_{1}\left[\epsilon\partial_{x}u-g_{1}  \right]\right),\\
u(0)&=f.
\end{split}
\end{equation}
The lifting operators $L_{i}\left(\cdot \right)$ in the SAT term \cite{arnold2002unified} are defined as $\int_{\Omega}\psi^{T}L_{i}\left(\phi \right)dx=\psi^{T}\phi\rvert_{x=x_{i}}$ for smooth scalar functions $\psi$ and $\phi$. The penalty parameters $\sigma_{0}$ and $\sigma_{1}$ will be determined to ensure energy boundedness. 

We again apply the energy method by multiplying (\ref{Eq:WeakImpositionBC_AdvectDiff}) with the solution vector, integrating by parts, and taking $\sigma_{0}=\sigma_{1}=-1$ to obtain
\begin{equation}
\label{Eq:AdvectDiff_ContEnergyEstWeakBC}
\dfrac{d}{dt}\norm{u}^{2} + 2\epsilon\norm{\partial_{x}u}^{2}=\dfrac{1}{a}\left.\left[g_{0}^{2}-(au-g_{0})^{2}\right]\right|_{x=0}+\dfrac{1}{a}\left.\left[g_{1}^{2}-(au-g_{1})^{2}\right]\right|_{x=0}\leq \dfrac{1}{a}\left(g_{0}^{2}+g_{1}^{2}\right),
\end{equation}
which is identical to (\ref{Eq:AdvectDiff_ContEnergyRate}). Time integration of (\ref{Eq:AdvectDiff_ContEnergyEstWeakBC}) leads to (\ref{Eq:AdvectDiff_ContEnergyEstStrongBC}). 

\section{The semi-discrete approximation}
\label{Sec:Semidiscrete_CGFEM}
For semi-discrete approximations in the CG framework, the computational domain $\Omega$ is discretized into $N_{\textrm{el}}$ finite elements such that $\Omega = \cup~\Omega^{e}$. The total number of global nodes becomes $(N+1)$, representing the number of nodes in the spatial direction, where $N= \left(N_{\textrm{el}}\times p\right)$ and $p$ refers to the degree of the Lagrange basis used.
We begin by constructing the spatial operators in SBP form at the element level~\cite{malan2023sbp}. These element-level operators are then assembled to form the global operators. This construction will be applied also for the time discretization. 
\subsection{Constructions of the SBP operators at the element level for one-dimensional domain}
\label{Sec:Construction}
 For a single element, we employ the Lagrange basis $(\mathcal{L}_{i})$ to approximate the solution. As such, the trial function $u(x,t)$ is approximated by:
 \begin{equation}
u^{h}(x,t) = \sum_{i=0}^{p}\mathcal{L}_{i}(x)U_{i}(t)=\bm{\mathcal{L}}^{T}(x)\bm{U}(t),
\end{equation}
where the total number of nodes per element is $\left(p+1\right)$, and $\bm{\mathcal{L}}(x)^{T}=[\mathcal{L}_{0}(x),\mathcal{L}_{2}(x),\ldots,$ $\mathcal{L}_{p}(x)]$ is the set of Lagrange polynomials of degree $p$. $\bm{U}(t)=[U_{0}(t),U_{1}(t),\ldots,U_{p}(t)]$ denotes the set of time-dependent coefficients at node points. 
 
We compute the element mass matrix $(\mathbf{P}^{e})$ and weak first order derivative operator $(\mathbf{Q}_{x}^{e})$ at the element level as (for details see \cite{malan2023sbp}): 
\begin{equation}
\label{Eq:SBP_OPS_1DM01}
\mathbf{P}^{e} = \int_{\Omega^{e}}\bm{\mathcal{L}}\bm{\mathcal{L}}^{T}~d\Omega^{e} \quad \textrm{and}\quad \mathbf{Q}_{x}^{e}=\int_{\Omega^{e}}\bm{\mathcal{L}}\left(\partial_{x}\bm{\mathcal{L}}\right)^{T}~d\Omega^{e},
\end{equation}
where $\int_{\Omega^{e}}(\cdot)~d\Omega^{e}$ defines the integral over the element domain. 
The operator $\mathbf{Q}_{x}^{e}$ is almost skew-symmetric since
\begin{equation}
\label{Eq:SBP_PropQx}
\mathbf{Q}_{x}^{e} = \int_{\Omega^{e}}\bm{\mathcal{L}}\left(\partial_{x}\bm{\mathcal{L}}\right)^{T} d\Omega^{e}= \left. \bm{\mathcal{L}} \bm{\mathcal{L}}^T \right|_{0}^{p}
-\int_{\Omega^{e}}\left(\partial_{x}\bm{\mathcal{L}}\right)\bm{\mathcal{L}}^{T}d\Omega^{e}=\tilde{\mathbf{B}}^{e}-\left({\mathbf{Q}}_{x}^{e}\right)^{T}, 
\end{equation}
where the boundary operator $\tilde{\mathbf{B}}^{e}=diag(-1,0,\ldots,0,1)$ contains non-zero values only on the boundary.

\noindent Similarly, we define the weak second-order derivative operator $\mathbf{Q}_{xx}^{e}$ in SBP form at the element level as:
\begin{equation}
\label{Eq:QxxSBP_OP_MM03}
\mathbf{Q}_{xx}^{e}=\int_{\Omega^{e}}\bm{\mathcal{L}}\left(\partial_{xx}\bm{\mathcal{L}}\right)^{T}d\Omega^{e}= \left.\bm{\mathcal{L}}\left(\partial_{x}\bm{\mathcal{L}}\right)^{T} \right|_{0}^{p} - \int_{\Omega^{e}}\left(\partial_{x}\bm{\mathcal{L}} \right)\left(\partial_{x}\bm{\mathcal{L}} \right)^{T}d\Omega^{e}.
\end{equation}

Next, we map each element from the physical domain $x\in [0,l_{x}^{e}]$ (where $l_{x}^{e}$ defines the length of the element) to the parametric domain $\xi\in[-1,1]$ using the Jacobian $J=\partial x/\partial \xi=l_{x}^{e}/2$ and compute the spatial integrals using Gauss-Lobatto (GL) quadrature. E.g., for a given function $f(x)$ the integral is computed as:
\begin{equation}
\int_{-1}^{1}f(x)dx=\sum_{i=0}^{p}f(\xi_{i})\omega_{i}|J|.
\end{equation}
Here $\xi_{i}$ refers to the GL quadrature nodes, and the corresponding quadrature weights are given by $\omega_{i}$. Hence, we compute $\mathbf{P}^{e}$ and $\mathbf{Q}_{x}^{e}$ in (\ref{Eq:SBP_OPS_1DM01}) as:
\begin{equation}
\begin{split}
\mathbf{P}^{e}&\approx \sum_{i=0}^{p} \bm{\mathcal{L}}(\xi_{i})\bm{\mathcal{L}}^{T}(\xi_{i})\omega_{i}|J| = |J|~\textrm{diag}[\omega_{0},\omega_{2},\ldots, \omega_{p}],\\
\mathbf{Q}_{x}^{e}&\approx\sum_{i=0}^{p}\bm{\mathcal{L}}(\xi_{i})\left(\partial_{x}\bm{\mathcal{L}}(\xi_{i})\right)^{T}\omega_{i}|J|. 
\end{split}
\end{equation}
\noindent Once we have $\mathbf{P}^{e}$ and $\mathbf{Q}_{x}^{e}$, we define the strong-form elemental first derivative SBP operator as:
\begin{equation}
\label{Eq:SBP_1stOrderOp}
\mathbf{D}_{x}^{e} = \left(\mathbf{P}^{e}\right)^{-1}\mathbf{Q}_{x}^{e}.
\end{equation}
 The weak form second-order derivative $\mathbf{Q}_{xx}^{e}$ in (\ref{Eq:QxxSBP_OP_MM03}) may now be formulated as:
\begin{equation}
\label{Eq:QxxSBP_OP_MM04}
\begin{split}
\mathbf{Q}_{xx}^{e}=\left.\bm{\mathcal{L}}\left(\mathbf{D}_{x}^{e}\bm{\mathcal{L}} \right)^{T}\right|_{0}^{p}- \left(\mathbf{D}_{x}^{e} \right)^{T}\mathbf{P}^{e}\mathbf{D}_{x}^{e}&=diag \left(-\left.\mathbf{D}_{x}^{e}\right|_{0},0,\ldots,0,\left.\mathbf{D}_{x}^{e}\right|_{p}\right)-\left(\mathbf{D}_{x}^{e} \right)^{T}\mathbf{P}^{e}\mathbf{D}_{x}^{e}\\
&=\tilde{\mathbf{B}}^{e}\mathbf{D}_{x}^{e} - \left(\mathbf{D}_{x}^{e} \right)^{T}\mathbf{P}^{e}\mathbf{D}_{x}^{e}.
\end{split}
\end{equation}
\noindent The corresponding strong-form second derivative SBP operator at the element level is:
\begin{equation}
\mathbf{D}_{xx}^{e}=\left(\mathbf{P}^{e}\right)^{-1}\mathbf{Q}_{xx}^{e}.
\end{equation}
\subsection{Construction of the global matrices}
\label{Sec:GlobalMatrixConstruction}
The single element matrices computed above are next assembled to obtain the global matrices, i.e., the global mass matrix $(\mathbf{P})$, the global $\mathbf{Q}_{x}$ matrix and the global $\mathbf{Q}_{xx}$ matrix. This process is exemplified by considering a two-element mesh. The global matrix $\mathbf{K}$ is merged as:
\begin{equation}
\mathbf{K} = \sum_{e=1}^{N_{\textrm{el}}} \mathbf{K}^{e}
=\begin{pmatrix} {K}^{L}_{00} & \ldots & {K}^{L}_{0p} & &  0 \\
\vdots & \ddots & \vdots & & \\
 K^{L}_{p0}&\ldots &\left( K^{L}_{pp} + K^{R}_{00}\right)& \ldots & K^{R}_{0p}\\
 & & \vdots&\ddots&\vdots \\
 0& & K^{R}_{p0} & \ldots &K^{R}_{pp}
 \end{pmatrix}.
\end{equation}
where the superscripts $L$ and $R$ denote the left and right element matrices, and the subscript is the associated row and column indices. The global mass matrix $\mathbf{P}$ is thus obtained for two element as:
\begin{equation}
\mathbf{P}=\sum_{e=1}^{N_{\textrm{el}}} \mathbf{P}^{e}=|J|~diag(\omega_{0},\omega_{1},\ldots,\omega_{p}+\omega_{0},\omega_{1},\ldots,\omega_{p}).
\end{equation}
The global $\mathbf{Q}_{x}$ matrix can now be obtained as
\begin{equation}
\begin{split}
\mathbf{Q}_{x}=\sum_{e=1}^{N_{\textrm{el}}}\mathbf{Q}_{x}^{e}&=\begin{pmatrix} {{Q}^{L}_{x}}_{00} & \ldots & {{Q}^{L}_{x}}_{0p} & &  0 \\
\vdots & \ddots & \vdots & & \\
 {{Q}^{L}_{x}}_{p0}&\ldots &\left( {{Q}^{L}_{x}}_{pp} + {{Q}^{R}_{x}}_{00}\right)& \ldots & {{Q}^{R}_{x}}_{0p}\\
 & & \vdots&\ddots&\vdots \\
 0& & {{Q}^{R}_{x}}_{p0} & \ldots &{{Q}^{R}_{x}}_{pp}
 \end{pmatrix}\\
 &=\begin{pmatrix} {{Q}^{L}_{x}}_{00} & \ldots & {{Q}^{L}_{x}}_{0p} & &  0 \\
\vdots & \ddots & \vdots & & \\
 {{Q}^{L}_{x}}_{p0}&\ldots &0& \ldots & {{Q}^{R}_{x}}_{0p}\\
 & & \vdots&\ddots&\vdots \\
 0& & {{Q}^{R}_{x}}_{p0} & \ldots &{{Q}^{R}_{x}}_{pp}
 \end{pmatrix},
 \end{split}
\end{equation}
where at the shared node ${{Q}^{L}_{x}}_{pp}=-{{Q}^{R}_{x}}_{00}$. The global $\mathbf{Q}_{xx}$ can now be computed as:
\begin{equation}
\begin{split}
\mathbf{Q}_{xx}=\sum_{e=1}^{N_{\textrm{el}}}\mathbf{Q}_{xx}^{e}&=diag \left(-\left.\mathbf{D}_{x}^{L}\right|_{0},0,\ldots,0,\left.\mathbf{D}_{x}^{L}\right|_{p}-\left.\mathbf{D}_{x}^{R}\right|_{0},0,\ldots,0,\left.\mathbf{D}_{x}^{R}\right|_{p}\right)\\
&-\biggl[\left(\mathbf{D}_{x}^{L} \right)^{T}\mathbf{P}^{L}\mathbf{D}_{x}^{L} + \left(\mathbf{D}_{x}^{R} \right)^{T}\mathbf{P}^{R}\mathbf{D}_{x}^{R}\biggr]\\
&=diag \left(-\left.\mathbf{D}_{x}^{L}\right|_{0},0,\ldots,0,0,0,\ldots,0,\left.\mathbf{D}_{x}^{R}\right|_{p}\right)\\
&-\biggl[\left(\mathbf{D}_{x}^{L} \right)^{T}\mathbf{P}^{L}\mathbf{D}_{x}^{L} + \left(\mathbf{D}_{x}^{R} \right)^{T}\mathbf{P}^{R}\mathbf{D}_{x}^{R}\biggr],
\end{split}
\end{equation}
at the shared node $\left.\mathbf{D}_{x}^{L}\bm{U}^{L}\right|_{p}\approx\left.\mathbf{D}_{x}^{R}\bm{U}^{R}\right|_{0}$, which is imposed strongly by removing $\mathbf{D}_{x}^{L}$ and $\mathbf{D}_{x}^{R}$ . After computing the global mass matrix $\mathbf{P}$, the global $\mathbf{Q}_{x}$, and the global $\mathbf{Q}_{xx}$, the global strong form SBP operators are computed as $\mathbf{D}_{x}=\mathbf{P}^{-1}\mathbf{Q}_{x}$ and $\mathbf{D}_{xx}=\mathbf{P}^{-1}\mathbf{Q}_{xx}$, respectively. 
\subsection{The semi-discrete energy analysis}
\label{Sec:SemiDiscrete_System}
We may now express the semi-discrete approximation of (\ref{Eq:WeakImpositionBC_AdvectDiff}) in the following form:
\begin{equation}
\label{Eq:SemiDiscreteAdvectionDiffProb}
\begin{split}
\mathbf{P}\partial_{t}\bm{U} + a\mathbf{Q}_{x}\bm{U} - \epsilon \mathbf{Q}_{xx}\bm{U} &= \mathbf{E}_{0}\bigl[\sigma_{0}\left(a\bm{U}-\epsilon \mathbf{D}_{x}\bm{U}-\bm{G}  \right)\bigr]+ \mathbf{E}_{N}\bigl[\sigma_{1}\left(\epsilon \mathbf{D}_{x}\bm{U}-\bm{G}  \right)\bigr],\\
\bm{U}(0) &= \bm{F},
\end{split}
\end{equation}
where $\mathbf{E}_{0}=\textrm{diag}(1,0,\ldots,0)$ and $\mathbf{E}_{N}=\textrm{diag}(0,\ldots,0,1)$.
We now obtain the energy estimate for the semi-discrete problem by multiplying (\ref{Eq:SemiDiscreteAdvectionDiffProb}) with $\bm{U}^{T}$ from the left and adding it to its transpose with $\sigma_{0}=\sigma_{N}=-1$, to obtain
\begin{equation}
\label{Eq:SemiDiscreteEnergyRate_AdvectDiff}
\dfrac{d}{dt}\norm{\bm{U}}^{2}_{\mathbf{P}} + 2\epsilon\norm{\mathbf{D}_{x}\bm{U}}^{2}_{\mathbf{P}}=\dfrac{1}{a}\left[\bm{G}_{0}^{2} -\left(a\bm{U}_{0}-\bm{G}_{0} \right)^{2} \right] + \dfrac{1}{a}\left[\bm{G}_{N}^{2} -\left(a\bm{U}_{N}-\bm{G}_{N} \right)^{2} \right],
\end{equation}
where $\norm{\bm{U}}^{2}_{\mathbf{P}}=\bm{U}^{T}\mathbf{P}\bm{U}$ and $\norm{\mathbf{D}_{x}\bm{U}}^{2}_{P}=\left(\mathbf{D}_{x}\bm{U}\right)^{T}\mathbf{P}\left(\mathbf{D}_{x}\bm{U} \right) $. The semi-discrete energy rate (\ref{Eq:SemiDiscreteEnergyRate_AdvectDiff}) mimics it's continuous counterpart term by term. Time integration of (\ref{Eq:SemiDiscreteEnergyRate_AdvectDiff}) leads to the energy estimate:
\begin{equation}
\label{Eq:EnergyEstimateSemiDiscreteSystem}
\begin{split}
\norm{\bm{U}(:,T)}^{2}_{\mathbf{P}} &+2\epsilon \int_{0}^{T}\norm{\mathbf{D}_{x}\bm{U}}^{2}_{\mathbf{P}}~dt\\&=\norm{\bm{F}_{0}}^{2}_{\mathbf{P}} + \dfrac{1}{a} \int_{0}^{T}\biggl[\left(\bm{G}_{0}^{2}+\bm{G}_{N}^{2}\right)-\left[\left(a\bm{U}_{0}-\bm{G}_{0} \right)^{2}-\left(a\bm{U}_{N}-\bm{G}_{N} \right)^{2} \right]\biggr]~dt,
\end{split}
\end{equation}
 which again mimics the continuous estimate (\ref{Eq:AdvectDiff_ContEnergyEstStrongBC}) term by term. 
\section{The fully-discrete system }
\label{Sec:FullyDiscrete_approximation}
We now discretize the temporal domain into $M_{\textrm{el}}$ number of finite elements to obtain the fully-discrete system. In addition to the spatial discretization, the temporal discretization generates $(M+1)\times (N+1)$ global nodes, where $M=(M_{\textrm{el}}\times p_{t})$. $p_{t}$ denotes the degree of the polynomial used in temporal-direction. This is the same procedure as in the spatial case discussed above. This results in the following vectors:
\begin{equation}
\begin{split}
\bm{U}&=\left(\bm{U}_{0}, \bm{U}_{1},\ldots,\bm{U}_{M} \right)^{T}, \qquad\qquad \bm{U}_{i} = \left(U_{i,0},U_{i,1},\ldots, U_{i,N} \right)^{T},\\
\bm{F}& = \left(\bm{F}_{0},\bm{U}_{1},\ldots,\bm{U}_{M} \right)^{T},\qquad\qquad \bm{F}_{0} = \left(f_{0},f_{1},f_{2},\ldots,f_{N} \right)^{T},\\
\bm{G}&=\left(\bm{G}_{0},\bm{G}_{1},\ldots,\bm{G}_{M} \right)^{T}, \ \qquad\qquad \bm{G}_{i}=\left(g_{0}(i\Delta t),0,0,\ldots,g_{1}(i\Delta t) \right)^{T},\\
 \bm{U}^{0}&=\left(U_{0,0}, U_{1,0},\ldots,U_{M,0} \right)^{T},\;\ \qquad \bm{G}^{0}=\left(g_{0}(0),g_{0}(\Delta t),\ldots,g_{0}(M\Delta t) \right)^{T},\\
\bm{U}^{N}&=\left(U_{0,N}, U_{1,N},\ldots,U_{M,N} \right)^{T}, \;\;\;\quad \bm{G}^{N}=\left(g_{N}(0),g_{N}(\Delta t),\ldots,g_{N}(M\Delta t) \right)^{T}.
\end{split}
\end{equation}
Considering the SBP operators and using the Kronecker product we now define the fully discrete system where the temporal and spatial grids are orthogonal to each other. The fully-discrete system is:
\begin{equation}
\label{Eq:FullyDiscreteSystem}
\begin{split}
\biggl[\left(\mathbf{Q}_{t}\otimes \mathbf{P}_{x} \right) +a\left(\mathbf{P}_{t}\otimes\mathbf{Q}_{x} \right) -\epsilon \left(\mathbf{P}_{t}\otimes \mathbf{Q}_{xx}  \right)  \biggr]\bm{U} &=\sigma_{t0}\left(\mathbf{E}_{t}\otimes \mathbf{P}_{x}  \right)(\bm{U}-\bm{F}) \\
&+ \sigma_{x0}\left(\mathbf{P}_{t}\otimes \mathbf{E}_{x0} \right)\left(a\bm{U}-\epsilon\mathbf{D}_{x}\bm{U}-\bm{G} \right)\\
&+\sigma_{x1}\left( \mathbf{P}_{t}\otimes \mathbf{E}_{xN}\right)\left(\epsilon \mathbf{D}_{x}\bm{U}-\bm{G} \right),
\end{split}
\end{equation}
where the subscripts define the associated matrix computed in that corresponding direction. $\mathbf{E}_{x0}$ and $\mathbf{E}_{xN}$ are identical to $\mathbf{E}_{0}$ and $\mathbf{E}_{N}$ in (\ref{Eq:SemiDiscreteAdvectionDiffProb}). The penalty term in the temporal direction $\sigma_{t0}$ will now be evaluated by performing the fully discrete energy stability analysis. 
\subsection{The fully discrete energy analysis}
\label{sec:FullyDiscreteEnergyAnalysis}
The energy estimate for the fully discrete system can be obtained by multiplying  (\ref{Eq:FullyDiscreteSystem}) with $\bm{U}^{T}$ from the left, and adding to its transpose with the choice $\sigma_{t0}=-1$, we obtain the following energy estimate:
\begin{equation}
\label{Eq:EnergyEstimateFullyDiscreteSystem}
\begin{split}
\bm{U}_{M}^{T}\mathbf{P}_{x}\bm{U}_{M} &+ 2\epsilon \left(\mathbf{D}_{x}\bm{U} \right)^{T}\left(\mathbf{P}_{t}\otimes \mathbf{P}_{x} \right)\left(\mathbf{D}_{x}\bm{U} \right)\\
&= \bm{F}_{0}^{T}\mathbf{P}_{x}\bm{F}_{0} +\dfrac{1}{a}\Biggl[\left(\bm{G}^{0}\right)^{T}\mathbf{P}_{t}\bm{G}^{0}+\left(\bm{G}^{1}\right)^{T}\mathbf{P}_{t}\bm{G}^{1} \\
&-\left[ \left( a\bm{U}^{0}-\bm{G}^{0}\right)^{T}\mathbf{P}_{t}\left(a\bm{U}^{0}-\bm{G}^{0} \right) + \left( a\bm{U}^{N}-\bm{G}^{N}\right)^{T}\mathbf{P}_{t}\left(a\bm{U}^{N}-\bm{G}^{N} \right) \right]\Biggr] \\
&-\left(\bm{U}_{0}-\bm{F}_{0} \right)^{T}\mathbf{P}_{x}\left(\bm{U}_{0}-\bm{F}_{0} \right).
\end{split}
\end{equation}
The energy estimate for the fully discrete system  (\ref{Eq:EnergyEstimateFullyDiscreteSystem}) is similar to the energy estimate of the continuous system (\ref{Eq:AdvectDiff_ContEnergyEstStrongBC}) and to the semi-discrete system (\ref{Eq:EnergyEstimateSemiDiscreteSystem}) with an additional damping term $-\left(\bm{U}_{0}-\bm{F}_{0} \right)^{T}\mathbf{P}_{x}\left(\bm{U}_{0}-\bm{F}_{0} \right)$. The estimate (\ref{Eq:EnergyEstimateFullyDiscreteSystem}) proves fully discrete energy stability.
%
\section{The multistage-based temporal formulation}
\label{Sec:MultiStage_TimeIntegration}
Alternatively, a one-step, multi-stage method can be employed using the multi-block technique \cite{lundquist2014sbp} to discretize the temporal domain. In this approach, the problem is solved successively over small time intervals using fewer grid points, where the solution at the end of each interval serves as the initial condition for the subsequent one. The obvious advantage of the multistage formulation is that a smaller system of equations can be solved at each stage. We start by discretizing the temporal domain into $M_{\textrm{el}}$ using the Lagrange polynomial of degree $p_{t}$, which eventually generates $(M+1)$ nodes in the time domain, where $M=\left(M_{\textrm{el}}\times p_{t} \right)$. Then, the corresponding time and solution vectors become $\bm{t}=(0=t_{0}, t_{1},t_{2},\ldots,t_{M}=T )^{T}$ and $\bm{U}=(U_{0},U_{1},U_{2},\ldots,U_{M})^{T}$, respectively. The original problem is now partitioned into $M$ subproblems, and we solve the subproblems subsequently, one after another.
One may write the discrete version of problem for the $i^{\textrm{th}}$ stage as:
\begin{equation}
\label{Eq:FullyDiscreteForm_Multistage}
\begin{split}
\biggl[\bigl(\mathbf{Q}_{{t,i}}&\otimes \mathbf{P}_{x} \bigr) +a\left(\mathbf{P}_{t,i}\otimes\mathbf{Q}_{x} \right) -\epsilon \left(\mathbf{P}_{t,i}\otimes\mathbf{Q}_{xx}  \right)  \biggr]\bm{U}_{i} =\left(\sigma_{t0}\mathbf{E}_{t,i}\otimes \mathbf{P}_{x}  \right)(\bm{U}_{i}-\bm{U}_{i-1}) \\
&+ \sigma_{x0}\left(\mathbf{P}_{t,i}\otimes\mathbf{E}_{x0} \right)\left(a\bm{U}-\epsilon\mathbf{D}_{x}\bm{U}-\bm{G}^{0} \right)+\sigma_{x1}\left( \mathbf{P}_{t,i}\otimes \mathbf{E}_{xN}\right)\left(\epsilon \mathbf{D}_{x}\bm{U}-\bm{G}^{N} \right).
\end{split}
\end{equation}
This equation will be solved to obtain the solution vector $\bm{U}_{i}=(\bm{U}_{i,0},\bm{U}_{i,1},\ldots,\bm{U}_{i,m_{i}})^{T}$.
\subsection{The fully discrete energy analysis for multistage based formulation}
\label{Sec:FullyDscreteEnergyAnalysis_Mutistage}
The associated energy estimate can now be obtained by multiplying (\ref{Eq:FullyDiscreteForm_Multistage}) with $\bm{U}_{i}^{T}$ from the left and adding to its transpose with the choices of $\sigma_{t0}=-1$ as follows:
\begin{equation}
\label{Eq:EnergyEstimateFullyDiscreteMultistage}
\begin{split}
\bm{U}_{i,m_{i}}^{T}\mathbf{P}_{x}&\bm{U}_{i,m_{i}} + 2\epsilon\left(\mathbf{D}_{x}\bm{U}_{i} \right)^{T}\left(\mathbf{P}_{t,i}\otimes\mathbf{P}_{x} \right)\left(\mathbf{D}_{x}\bm{U}_{i} \right)\\&=\bm{U}_{i-1,0}^{T}\mathbf{P}_{x}\bm{U}_{i-1,0}+\dfrac{1}{a}\Biggl[\left(\bm{G}^{0}\right)^{T}\mathbf{P}_{t}\bm{G}^{0}+\left(\bm{G}^{N}\right)^{T}\mathbf{P}_{t,i}\bm{G}^{N} \\
&-\left[ \left( a\bm{U}^{0}-\bm{G}^{0}\right)^{T}\mathbf{P}_{t,i}\left(a\bm{U}^{0}-\bm{G}^{0} \right) + \left( a\bm{U}^{N}-\bm{G}^{N}\right)^{T}\mathbf{P}_{t,i}\left(a\bm{U}^{N}-\bm{G}^{N} \right) \right]\Biggr] \\
&-\left(\bm{U}_{i,0}-\bm{U}_{i-1,0} \right)^{T}\mathbf{P}_{x}\left(\bm{U}_{i,0}-\bm{U}_{i-1,0} \right).
\end{split}
\end{equation}
The energy estimate in (\ref{Eq:EnergyEstimateFullyDiscreteMultistage}) is similar to that of the global system in (\ref{Eq:EnergyEstimateFullyDiscreteSystem}) and proves energy stability of the fully discrete scheme for the multistage based formulation  (\ref{Eq:FullyDiscreteForm_Multistage}). 
\section{Numerical results}
\label{Sec:Numerical_results_advectDiff_results}
To analyze the performance of the new scheme, we perform mesh convergence studies using the MMS technique\cite{roache2002code}. To evaluate the mesh convergence in time and mesh separately, we consider the following analytical solution: 
\begin{equation}
\label{Eq:TemporalConv_EqAD}
u(x,t)=\textrm{exp}\left[\frac{x-1}{\epsilon_{1}}\right]\sin(x-at)\cos\left(\frac{\omega}{\epsilon_{2}}t\right),
\end{equation}
where $\omega=2\pi$ and $\epsilon_{1},\epsilon_{2}$ take various values depending on the case being studied.

\subsection{MMS study: temporal mesh convergence}
 We have solved this problem for three distinct cases where $\epsilon_{1}=1$ and $\epsilon_{2}$ takes the values $1,1/4$ and $1/8$. To assess the temporal order of convergence of the fully discrete formulation, sufficiently fine spatial grids are employed to ensure that the spatial discretization errors are negligible. We have considered a Lagrange polynomial of degree $p_x=8$ within each element, supported over 100 elements, yielding $N_x=801$ number of spatial nodes. Both the global and the multistage technique are employed to discretize the problem in the temporal direction. In the multistage technique, the number of nodes per stage is $(p_{t}+1)$. Symbols $N_{t}$ and $p_{t}$ denote the number of nodes and degree of the polynomials used in temporal direction, respectively. The convergence studies are provided in Tables~\ref{tab:epsilon1_1_adD_temporalConv_epsilon2_1b1_1b4}-\ref{tab:epsilon1_1_adD_temporalConv_epsilon2_1b8} and show super-convergence behavior for $p_t\geq 2$, with an accuracy $\mathcal{O}(p_{t}+2)$. The linear basis provides an accuracy of $\mathcal{O}(p_{t}+1)$. 
\begin{remark}
 \label{Remark_1}
The global and multistage methods yield the same order of convergence, but the multistage-based formulation significantly reduces the error compared to the global method.
\end{remark}

\begin{table*}[ht!]
\centering
\caption{Temporal convergence studies}
\label{tab:epsilon1_1_adD_temporalConv_epsilon2_1b1_1b4}
\resizebox{\linewidth}{!}{%
\begin{tabular}{cccccccccccc}
\hline
\multicolumn{12}{c}{$a=1,\epsilon=0.01,\epsilon_{1}=1$ and $\epsilon_{2}=1$ } \\ \hline
\multicolumn{6}{c}{$p_{t}=1$} & \multicolumn{6}{c}{$p_{t}=2$} \\ \hline
\multirow{2}{*}{$N_{t}$}   & \multirow{2}{*}{$CFL$} & \multicolumn{2}{c}{Global} 	 & \multicolumn{2}{c}{Multistage}     & \multirow{2}{*}{$N_{t}$} & \multirow{2}{*}{$CFL$} & \multicolumn{2}{c}{Global} & \multicolumn{2}{c}{Multistage} \\ 
\cmidrule(lr){3-6} \cmidrule(lr){9-12} 
  			   &          		   & $\norm{E}_{Nt}$  & $\mathcal{O}(E)$ & $\norm{E}_{Nt}$ & $\mathcal{O}(E)$ &    			 & 	  		 & $\norm{E}_{Nt}$ & $\mathcal{O}(E)$ & $\norm{E}_{Nt}$ & $\mathcal{O}(E)$ \\ \hline
3  			   & 997.586 		  &5.720e+00          &-                &6.213e+00         &-                 &5     			 & 498.793   		 &1.360e+00          &-               &6.904e-02        &-        \\
5  			   & 498.793 		  &1.550e+00          &   2.556         &1.515e+00         &   2.763	      &9     			 & 249.396   		 &7.235e-02          &   4.992        &4.753e-03        &   4.552 \\
7  			   & 332.529 		  &7.703e-01          &   2.079         &6.891e-01         &   2.341	      &13    			 & 166.264   		 &1.224e-02          &   4.832        &9.934e-04        &   4.257 \\
9  			   & 249.396 		  &4.906e-01          &   1.795         &4.044e-01         &   2.120	      &17    			 & 124.698   		 &3.675e-03          &   4.485        &3.260e-04        &   4.154 \\
11 			   & 199.517 		  &3.276e-01          &   2.012         &2.692e-01         &   2.027	      &21    			 &  99.759   		 &1.477e-03          &   4.314        &1.373e-04        &   4.093 \\
13 			   & 166.264 		  &2.280e-01          &   2.170         &1.929e-01         &   1.998	      &25    			 &  83.132   		 &7.085e-04          &   4.213        &6.770e-05        &   4.053 \\
15 			   & 142.512 		  &1.670e-01          &   2.176         &1.450e-01         &   1.993	      &29    			 &  71.256   		 &3.829e-04          &   4.147        &3.726e-05        &   4.024\\ \hline
\multicolumn{6}{c}{$p_{t}=3$} & \multicolumn{6}{c}{$p_{t}=4$} \\ \hline
\multirow{2}{*}{$N_{t}$} & \multirow{2}{*}{$CFL$} & \multicolumn{2}{c}{Global} & \multicolumn{2}{c}{Multistage}               & \multirow{2}{*}{$N_{t}$} & \multirow{2}{*}{$CFL$}& \multicolumn{2}{c}{Global} & \multicolumn{2}{c}{Multistage} \\ 
\cmidrule(lr){3-6} \cmidrule(lr){9-12} 
  			 &                        & $\norm{E}_{Nt}$   & $\mathcal{O}(E)$ & $\norm{E}_{Nt}$ & $\mathcal{O}(E)$ &                          &                      & $\norm{E}_{Nt}$   & $\mathcal{O}(E)$ & $\norm{E}_{Nt}$ & $\mathcal{O}(E)$ \\ \hline
13  			 & 137.863 		  &4.756e-03          &   -             &2.710e-05         &   -	      &25    			 &  57.419    		&2.387e-05          &   6.836         &3.755e-09         &   7.233\\
19  			 &  91.909 		  &6.793e-04          &   5.128         &3.992e-06         &   5.047	      &33    			 &  43.064    		&3.668e-06          &   6.746         &5.421e-10         &   6.971\\
25  			 &  68.931 		  &1.672e-04          &   5.108         &9.643e-07         &   5.176	      &41    			 &  34.451    		&8.671e-07          &   6.644         &1.236e-10         &   6.810\\
31  			 &  55.145 		  &5.697e-05          &   5.006         &3.138e-07         &   5.219	      &49    			 &  28.709    		&2.698e-07          &   6.550         &3.756e-11         &   6.683\\
37  			 &  45.954 		  &2.366e-05          &   4.966         &1.250e-07         &   5.200	      &57    			 &  24.608    		&1.014e-07          &   6.470         &1.394e-11         &   6.557\\
43  			 &  39.389 		  &1.126e-05          &   4.942         &5.756e-08         &   5.162	      &61    			 &  22.968    		&6.562e-08          &   6.418         &9.025e-12         &   6.405\\
46  			 &  36.763 		  &8.076e-06          &   4.928         &4.072e-08         &   5.131	      &65    			 &  21.532    		&4.373e-08          &   6.387         &6.067e-12         &   6.252\\ \hline
\end{tabular}%
}
\end{table*}

\begin{table*}[ht!]
\centering
\caption{Temporal convergence studies}
\label{tab:epsilon1_1_adD_temporalConv_epsilon2_1b4}
\resizebox{1\textwidth}{!}{%
\begin{tabular}{cccccccccccc}
\hline
\multicolumn{12}{c}{$a=1,\epsilon=0.01,\epsilon_{1}=1$ and $\epsilon_{2}=1/4$ } \\ \hline
\multicolumn{6}{c}{$p_{t}=1$} & \multicolumn{6}{c}{$p_{t}=2$} \\ \hline
\multirow{2}{*}{$N_{t}$} & \multirow{2}{*}{$CFL$} & \multicolumn{2}{c}{Global} & \multicolumn{2}{c}{Multistage} & \multirow{2}{*}{$N_{t}$}                & \multirow{2}{*}{$CFL$} & \multicolumn{2}{c}{Global} & \multicolumn{2}{c}{Multistage} \\ 
\cmidrule(lr){3-6} \cmidrule(lr){9-12} 
                         &                        & $\norm{E}_{Nt}$   & $\mathcal{O}(E)$  & $\norm{E}_{Nt}$ & $\mathcal{O}(E)$ &                          &                      & $\norm{E}_{Nt}$  & $\mathcal{O}(E)$ & $\norm{E}_{Nt}$ & $\mathcal{O}(E)$ \\ \hline
17   			 & 124.698  		  &1.617e+00          &-                &1.590e+00          &-       	       &25   			  &  83.132     	 &8.723e-02         &   -              &4.951e-03        &   - \\
21   			 &  99.759  		  &1.037e+00          &   2.101         &1.008e+00          &   2.157	       &29   			  &  71.256     	 &4.279e-02         &   4.799          &2.692e-03        &   4.105 \\
25   			 &  83.132  		  &7.253e-01          &   2.053         &6.992e-01          &   2.099	       &33   			  &  62.349     	 &2.355e-02         &   4.622          &1.590e-03        &   4.074 \\
29   			 &  71.256  		  &5.356e-01          &   2.044         &5.146e-01          &   2.066	       &37   			  &  55.421     	 &1.408e-02         &   4.497          &1.000e-03        &   4.054 \\
33   			 &  62.349  		  &4.113e-01          &   2.043         &3.950e-01          &   2.046	       &41   			  &  49.879     	 &8.953e-03         &   4.408          &6.605e-04        &   4.041 \\
37   			 &  55.421  		  &3.256e-01          &   2.043         &3.130e-01          &   2.033	       &49   			  &  41.566     	 &4.147e-03         &   4.318          &3.221e-04        &   4.028 \\
41   			 &  49.879  		  &2.640e-01          &   2.043         &2.543e-01          &   2.025	       &53   			  &  38.369     	 &2.970e-03         &   4.253          &2.350e-04        &   4.019 \\ \hline
\multicolumn{6}{c}{$p_{t}=3$} & \multicolumn{6}{c}{$p_{t}=4$} \\ \hline
\multirow{2}{*}{$N_{t}$} & \multirow{2}{*}{$CFL$} & \multicolumn{2}{c}{Global} & \multicolumn{2}{c}{Multistage} & \multirow{2}{*}{$N_{t}$}                & \multirow{2}{*}{$CFL$} & \multicolumn{2}{c}{Global} & \multicolumn{2}{c}{Multistage} \\ 
\cmidrule(lr){3-6} \cmidrule(lr){9-12} 
                         &                         & $\norm{E}_{Nt}$   & $\mathcal{O}(E)$  & $\norm{E}_{Nt}$ &$\mathcal{O}(E)$&                           &                      & $\norm{E}_{Nt}$  & $\mathcal{O}(E)$  & $\norm{E}_{Nt}$ & $\mathcal{O}(E)$ \\ \hline
25    			 &  68.931    		   &2.987e-02          &   -               &1.422e-04        &   -           	&137    		  &  10.133     	 &2.704e-07          &   6.597         &2.872e-11         &   6.663 \\
31    			 &  55.145    		   &1.171e-02          &   4.355           &5.459e-05        &   4.450       	&153    		  &   9.066     	 &1.313e-07          &   6.537         &1.385e-11         &   6.604 \\
37    			 &  45.954    		   &5.148e-03          &   4.644           &2.441e-05        &   4.549       	&169    		  &   8.203     	 &6.893e-08          &   6.482         &7.210e-12         &   6.562 \\
43    			 &  39.389    		   &2.498e-03          &   4.812           &1.186e-05        &   4.802       	&185    		  &   7.489     	 &3.853e-08          &   6.432         &3.995e-12         &   6.526 \\
49    			 &  34.466    		   &1.321e-03          &   4.880           &6.257e-06        &   4.897       	&201    		  &   6.890     	 &2.268e-08          &   6.388         &2.336e-12         &   6.471 \\
55    			 &  30.636    		   &7.500e-04          &   4.899           &3.539e-06        &   4.935       	&217    		  &   6.380     	 &1.395e-08          &   6.349         &1.430e-12         &   6.411 \\
61    			 &  27.573    		   &4.515e-04          &   4.900           &2.121e-06        &   4.946       	&233    		  &   5.940     	 &8.899e-09          &   6.315         &9.184e-13         &   6.221 \\
67    			 &  25.066    		   &2.852e-04          &   4.896           &1.334e-06        &   4.944       	&249    		  &   5.557     	 &5.862e-09          &   6.286         &6.144e-13         &   6.052 \\ \hline
\end{tabular}%
}
\end{table*}
\begin{table*}[ht!]
\centering
\caption{Temporal convergence studies}
\label{tab:epsilon1_1_adD_temporalConv_epsilon2_1b8}
\resizebox{1\textwidth}{!}{%
\begin{tabular}{cccccccccccc}
\hline
\multicolumn{12}{c}{$a=1,\epsilon=0.01,\epsilon_{1}=1$ and $\epsilon_{2}=1/8$ } \\ \hline
\multicolumn{6}{c}{$p_{t}=1$} & \multicolumn{6}{c}{$p_{t}=2$} \\ \hline
\multirow{2}{*}{$N_{t}$}   & \multirow{2}{*}{$CFL$} & \multicolumn{2}{c}{Global} 	 & \multicolumn{2}{c}{Multistage}     & \multirow{2}{*}{$N_{t}$} & \multirow{2}{*}{$CFL$} & \multicolumn{2}{c}{Global} & \multicolumn{2}{c}{Multistage} \\ 
\cmidrule(lr){3-6} \cmidrule(lr){9-12} 
  			   &          		   & $\norm{E}_{Nt}$   & $\mathcal{O}(E)$& $\norm{E}_{Nt}$ & $\mathcal{O}(E)$ &    			 & 	  		 & $\norm{E}_{Nt}$   & $\mathcal{O}(E)$  & $\norm{E}_{Nt}$ & $\mathcal{O}(E)$ \\ \hline
37 			   &  55.421  		   &1.254e+00          &   -             &1.249e+00        &   -	      &33  			 &  62.349   		 &4.380e-01          &-               	 &1.972e-02        &-        \\
43 			   &  47.504  		   &9.161e-01          &   2.092         &9.103e-01        &   2.103	      &49  			 &  41.566   		 &6.658e-02          &   4.765        	 &3.755e-03        &   4.196 \\
49 			   &  41.566  		   &6.992e-01          &   2.068         &6.939e-01        &   2.079	      &65  			 &  31.175   		 &1.906e-02          &   4.427        	 &1.183e-03        &   4.087 \\
55 			   &  36.948  		   &5.515e-01          &   2.054         &5.468e-01        &   2.062	      &81  			 &  24.940   		 &7.445e-03          &   4.272        	 &4.858e-04        &   4.045 \\
61 			   &  33.253  		   &4.463e-01          &   2.045         & 4.422e-01       &   2.050	      &97  			 &  20.783   		 &3.498e-03          &   4.190        	 &2.352e-04        &   4.025 \\
73 			   &  27.711  		   &3.096e-01          &   2.036         &3.067e-01        &   2.039	      &113 			 &  17.814   		 &1.858e-03          &   4.142        	 &1.274e-04        &   4.014 \\
79 			   &  25.579  		   &2.638e-01          &   2.030         &2.612e-01        &   2.030	      &129 			 &  15.587   		 &1.078e-03          &   4.111        	 &7.494e-05        &   4.007 \\
85 			   &  23.752  		   &2.274e-01          &   2.027         &2.252e-01        &   2.026	      &145 			 &  13.855   		 &6.684e-04          &   4.090        	 &4.693e-05        &   4.003 \\ \hline
\multicolumn{6}{c}{$p_{t}=3$} & \multicolumn{6}{c}{$p_{t}=4$} \\ \hline
\multirow{2}{*}{$N_{t}$} & \multirow{2}{*}{$CFL$}  & \multicolumn{2}{c}{Global} & \multicolumn{2}{c}{Multistage}               & \multirow{2}{*}{$N_{t}$} & \multirow{2}{*}{$CFL$}& \multicolumn{2}{c}{Global} & \multicolumn{2}{c}{Multistage} \\ 
\cmidrule(lr){3-6} \cmidrule(lr){9-12} 
  			 &                         & $\norm{E}_{Nt}$   & $\mathcal{O}(E)$ & $\norm{E}_{Nt}$ & $\mathcal{O}(E)$ &                         &                      & $\norm{E}_{Nt}$   & $\mathcal{O}(E)$ & $\norm{E}_{Nt}$   & $\mathcal{O}(E)$ \\ \hline
55   			 &  30.636  		   &9.974e-03          &-                &4.404e-05         &-       	       &49    			 &  28.709   		&1.151e-02          &-                 &1.611e-06          &-       \\
109  			 &  15.318  		   &4.033e-04          &   4.690         &1.870e-06         &   4.618	       &101   			 &  13.781   		&9.831e-05          &   6.586          &1.029e-08          &   6.987\\
217  			 &   7.659  		   &1.525e-05          &   4.756         &7.120e-08         &   4.747	       &197   			 &   7.031   		&1.110e-06          &   6.711          &1.023e-10          &   6.902\\
325  			 &   5.106  		   &2.292e-06          &   4.692         &1.056e-08         &   4.724	       &245   			 &   5.648   		&2.646e-07          &   6.575          &2.359e-11          &   6.727\\
433  			 &   3.830  		   &6.020e-07          &   4.660         &2.751e-09         &   4.689	       &293   			 &   4.719   		&8.230e-08          &   6.528          &7.353e-12          &   6.515\\
541  			 &   3.064  		   &2.129e-07          &   4.668         &9.789e-10         &   4.639	       &341   			 &   4.053   		&3.068e-08          &   6.505          &2.802e-12          &   6.360\\
649  			 &   2.553  		   &9.074e-08          &   4.686         &4.213e-10         &   4.631	       &389   			 &   3.552   		&1.304e-08          &   6.497          &1.219e-12          &   6.318\\
703  			 &   2.357  		   &6.233e-08          &   4.697         &2.907e-10         &   4.644	       &437   			 &   3.161   		&6.124e-09          &   6.494          &5.984e-13          &   6.118\\
757  			 &   2.188  		   &4.401e-08          &   4.703         &2.060e-10         &   4.655	       &485   			 &   2.847   		&3.114e-09          &   6.489          &3.161e-13          &   6.122\\ \hline
\end{tabular}%
}
\end{table*}

\subsection{MMS study: spatial mesh convergence}
We next study the order of mesh convergence in the spatial direction and again consider the analytical solution (\ref{Eq:TemporalConv_EqAD}). The problem is solved for three distinct cases where $\epsilon_{2}=1$ and $\epsilon_{1}$ takes the values $1, 1/10$ and $1/100$. To determine the spatial order of convergence, sufficiently fine temporal grids are employed to ensure that the temporal errors are negligible. We have discretized the temporal domain using a Lagrange polynomial of degree $p_{t}=10$ within each element, supported over 200 elements, yielding 2001 nodes. The multistage technique is employed, with $(p_t+1)$ number of nodes per stage. For the spatial discretization, a series of different grids is used, where $N_{x}$ and $p_{x}$ define the number of nodes and degree of the Lagrange basis, respectively. The convergence studies are provided in  Table~\ref{tab:SpatialConv_Eps2_1_Eps1_1_1b100}. 
It shows super-convergence behavior for $p_x\geq 2$, with accuracy $\mathcal{O}(p_{x}+2)$. For linear polynomials, the order of accuracy is $\mathcal{O}(p_{x}+1)$. 

\begin{table*}[ht!]
\centering
\caption{Spatial convergence studies}
\label{tab:SpatialConv_Eps2_1_Eps1_1_1b100}
\resizebox{\textwidth}{!}{%
\begin{tabular}{cccccccccccc}
\hline
\multicolumn{12}{c}{$a=1,\epsilon=0.01,\epsilon_1=1,\epsilon_2=1$}  \\
\hline
 \multicolumn{3}{c}{$p_x=1$}       & \multicolumn{3}{c}{$p_x=2$}       &  \multicolumn{3}{c}{$p_x=3$}           &  \multicolumn{3}{c}{$p_x=4$}       \\ \cmidrule(lr){1-3} \cmidrule(lr){4-6} \cmidrule(lr){7-9} \cmidrule(lr){10-12} 
 $N_x$   & $\norm{E}_{Nt}$   & $\mathcal{O}(E)$ &   $N_x$ & $\norm{E}_{Nt}$           & $\mathcal{O}(E)$       &  $N_x$     & $\norm{E}_{Nt}$ & $\mathcal{O}(E)$&  $N_x$  & $\norm{E}_{Nt}$  & $\mathcal{O}(E)$ \\ \hline
21       &2.971e-04          &   -       	 &    25       &1.788e-05              &   -		         &49      &7.781e-09              &   -            &    65       &3.028e-11              &   -   \\
25       &1.902e-04          &   2.557       	 &    33       &5.901e-06              &   3.993 		 &61      &2.849e-09              &   4.587        &    81       &8.404e-12              &   5.825  \\
29       &1.316e-04          &   2.483       	 &    41       &2.456e-06              &   4.038 		 &73      &1.228e-09              &   4.684        &    97       &2.925e-12              &   5.854  \\
33       &9.633e-05          &   2.414       	 &    49       &1.194e-06              &   4.048 		 &85      &5.956e-10              &   4.757        &    113      &1.192e-12              &   5.881  \\
37       &7.358e-05          &   2.355       	 &    57       &6.471e-07              &   4.048 		 &97      &3.155e-10              &   4.810        &    129      &5.450e-13              &   5.908  \\
41       &5.808e-05          &   2.305       	 &    65       &3.804e-07              &   4.046 		 &109     &1.792e-10              &   4.850        &    145      &2.726e-13              &   5.924  \\
45       &4.704e-05          &   2.265       	 &    73       &2.379e-07              &   4.043 		 &121     &1.077e-10              &   4.880        &    161      &1.468e-13              &   5.915  \\
49       &3.890e-05          &   2.231       	 &    81       &1.563e-07              &   4.040 		 &133     &6.772e-11              &   4.902        &    177      &8.323e-14              &   5.989  \\
53       &3.272e-05          &   2.203       	 &    89       &1.068e-07              &   4.037 		 &145     &4.427e-11              &   4.920        &    193      &5.073e-14              &   5.720   \\ \hline
\multicolumn{12}{c}{$a=1,\epsilon=0.01,\epsilon_{1}=1/10,\epsilon_2=1$}  \\
\hline
 \multicolumn{3}{c}{$p_x=1$}       & \multicolumn{3}{c}{$p_x=2$}       &  \multicolumn{3}{c}{$p_x=3$}           &  \multicolumn{3}{c}{$p_x=4$}       \\ \cmidrule(lr){1-3} \cmidrule(lr){4-6} \cmidrule(lr){7-9} \cmidrule(lr){10-12} 
 $N_x$   & $\norm{E}_{Nt}$   & $\mathcal{O}(E)$ &   $N_x$ & $\norm{E}_{Nt}$           & $\mathcal{O}(E)$       &  $N_x$     & $\norm{E}_{Nt}$ & $\mathcal{O}(E)$&  $N_x$  & $\norm{E}_{Nt}$  & $\mathcal{O}(E)$ \\ \hline
29      &1.008e-03           &   -               &    33      &1.839e-04              &   -			  &49      &5.568e-06              &   -            &  81       &4.488e-08              &   -      \\
33      &7.140e-04           &   2.673           &    41      &7.992e-05              &   3.840			  &61      &2.019e-06              &   4.630        &  97       &1.569e-08              &   5.830 \\
37      &5.315e-04           &   2.580           &    49      &3.994e-05              &   3.892			  &73      &8.679e-07              &   4.702        &  113      &6.407e-09              &   5.866  \\
41      &4.111e-04           &   2.502           &    57      &2.206e-05              &   3.924			  &85      &4.204e-07              &   4.763        &  129      &2.936e-09              &   5.894  \\
45      &3.276e-04           &   2.437           &    65      &1.314e-05              &   3.944			  &97      &2.228e-07              &   4.810        &  145      &1.470e-09              &   5.915  \\
49      &2.675e-04           &   2.383           &    73      &8.301e-06              &   3.959			  &109     &1.266e-07              &   4.847        &  161      &7.901e-10              &   5.932  \\
53      &2.226e-04           &   2.338           &    81      &5.494e-06              &   3.969			  &121     &7.606e-08              &   4.876        &  177      &4.499e-10              &   5.944  \\
57      &1.883e-04           &   2.300           &    89      &3.778e-06              &   3.976			  &133     &4.787e-08              &   4.898        &  193      &2.687e-10              &   5.955  \\
61      &1.615e-04           &   2.268           &    97      &2.682e-06              &   3.982			  &145     &3.131e-08              &   4.915        &  209      &1.671e-10              &   5.963   \\ \hline
\multicolumn{12}{c}{$a=1,\epsilon=0.01,\epsilon_1=1/100,\epsilon_2=1$}  \\
\hline
 \multicolumn{3}{c}{$p_x=1$}       & \multicolumn{3}{c}{$p_x=2$}       &  \multicolumn{3}{c}{$p_x=3$}           &  \multicolumn{3}{c}{$p_x=4$}       \\ \cmidrule(lr){1-3} \cmidrule(lr){4-6} \cmidrule(lr){7-9} \cmidrule(lr){10-12} 
 $N_x$   & $\norm{E}_{Nt}$   & $\mathcal{O}(E)$ &   $N_x$ & $\norm{E}_{Nt}$           & $\mathcal{O}(E)$       &  $N_x$     & $\norm{E}_{Nt}$ & $\mathcal{O}(E)$&  $N_x$  & $\norm{E}_{Nt}$  & $\mathcal{O}(E)$ \\ \hline
31      &2.361e-03              &   -           &    97       &2.944e-05              &   3.188    		  &37       &3.839e-04              &-               &  49       &1.362e-04              &-        \\
37      &1.563e-03              &   2.330       &    121      &1.393e-05              &   3.384    		  &109      &1.115e-05              &   3.275        &  145      &1.600e-06              &   4.096 \\
43      &1.112e-03              &   2.267       &    145      &7.346e-06              &   3.537    		  &145      &3.452e-06              &   4.110        &  193      &3.663e-07              &   5.157  \\
49      &8.320e-04              &   2.220       &    169      &4.203e-06              &   3.646    		  &181      &1.307e-06              &   4.379        &  241      &1.098e-07              &   5.425  \\
55      &6.464e-04              &   2.185       &    193      &2.563e-06              &   3.723    		  &217      &5.726e-07              &   4.550        &  289      &3.979e-08              &   5.588  \\
61      &5.169e-04              &   2.158       &    217      &1.646e-06              &   3.780    		  &253      &2.800e-07              &   4.662        &  337      &1.659e-08              &   5.693  \\
67      &4.230e-04              &   2.137       &    241      &1.102e-06              &   3.821    		  &289      &1.490e-07              &   4.739        &  385      &7.700e-09              &   5.764  \\
73      &3.527e-04              &   2.120       &    265      &7.647e-07              &   3.852    		  &325      &8.490e-08              &   4.793        &  433      &3.889e-09              &   5.814  \\
79      &2.987e-04              &   2.106       &    289      &5.464e-07              &   3.877    		  &361      &5.110e-08              &   4.833        &  481      &2.103e-09              &   5.850   \\ 
85      &2.562e-04              &   2.094       &    313      &4.005e-07              &   3.896    		  &397      &3.219e-08              &   4.863        &  529      &1.202e-09              &   5.877   \\ \hline
\end{tabular}%
}
\end{table*}
\subsection{A typical application}
\label{Sec:NumericalProbBounInducedDiscontinuties}
Finally, we solve a combined wave propagation and boundary layer problem considering the fully-discrete space-time SBP method. The Robin boundary condition $au(0,t) - \epsilon u_x(0,t) = g(t)$ is imposed at the inflow boundary ($x=0$). First, we consider a steady-state problem with $g = 1$. Secondly, we solve a transient problem, where $g(t)$ takes the  form:
\begin{equation}
\label{Eq:InflowBC_timeVarying_g}
    g(t) = 
    \begin{cases}
        1, & t < 0.05,\\[4pt]
        1 + \alpha\sin\!\left(\omega (t - 0.05)\right), 
        & t \ge 0.05,
    \end{cases}
\end{equation}
with frequency $\omega=10\pi$ and amplitude $\alpha = 0.5$. At the outflow boundary ($x=1$), we have imposed $u=0$.

For the first problem, a deliberately incorrect initial solution $u(0,0)=0$ is considered and we study how fast we reach the correct steady solution. For the second problem, we start by imposing $u(0,0)=1$, and then the time varying wave is generated at $t\geq0.5$  following (\ref{Eq:InflowBC_timeVarying_g}). While solving this type of time-varying wave propagation problem, many schemes induce spurious oscillations unless a sufficiently fine grid is employed, especially when advection dominates diffusion as the wave propagates through the domain. 

Here, we have considered an advection dominated flow with $\epsilon=0.01$. We construct the reference solutions for each case using the explicit Runge-Kutta $4^{\textrm{th}}$-order (RK4) with CFL = 0.01 and a very fine spatial mesh consisting of $8^{\textrm{th}}$ degree Lagrange polynomial supported over $300$ elements. The reference solutions are mesh and time converged for all cases. 
\begin{figure*}[ht!]
		        \centering
		 \begin{subfigure}[h!]{0.5\textwidth}
        \centering
        \includegraphics[trim={0.0cm 0.0cm 0.0cm 0.0cm}, clip=true, width=1.0\textwidth]{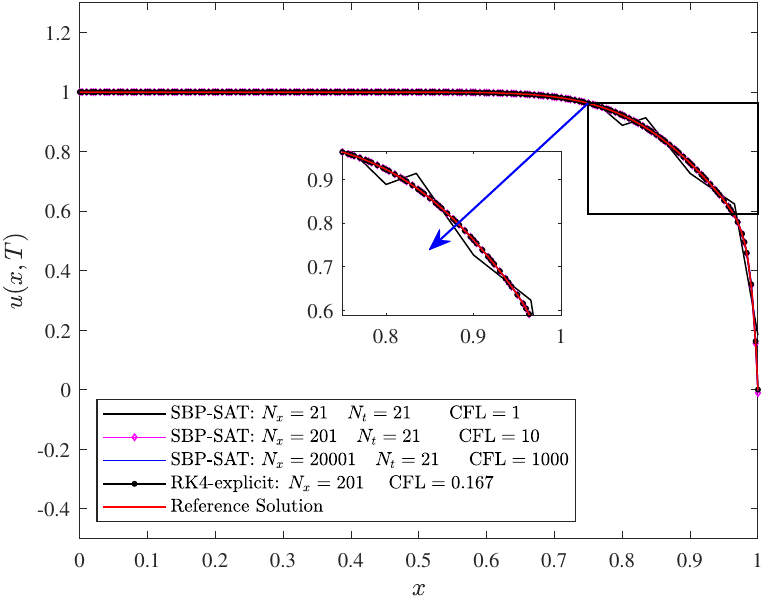}
\end{subfigure}%
 \begin{subfigure}[ht!]{0.5\textwidth}
		        \centering
		\includegraphics[trim={0.0cm 0.0cm 0.0cm 0.0cm}, clip=true, width=1.0\textwidth]{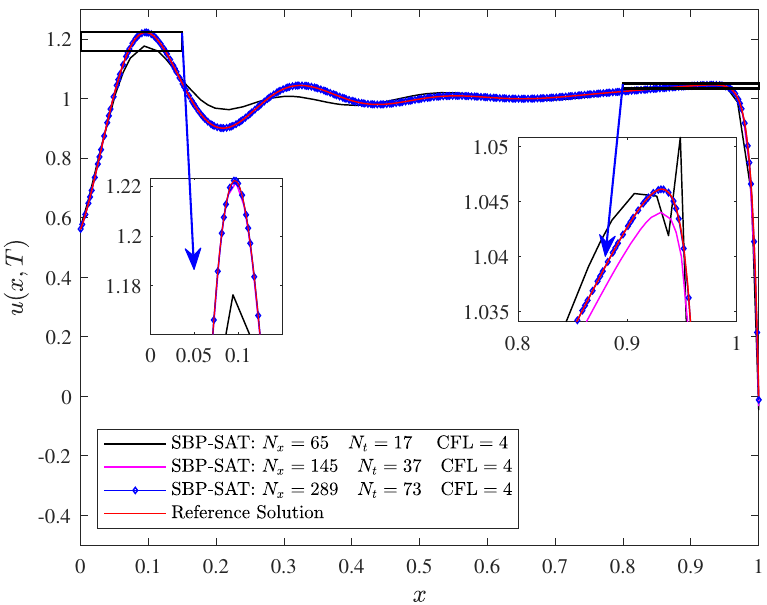}
\end{subfigure}
    \caption{Convergence of solution at time $t=1$ sec for (left) $g=1$ and (right) $g(t)$ takes the form of (\ref{Eq:InflowBC_timeVarying_g})}
\label{Fig:WavePropagation_SBP_spaceTime}
    \end{figure*} 
     \begin{figure*}[ht!]
		        \centering
		 \begin{subfigure}[h!]{0.5\textwidth}
        \centering
        \includegraphics[trim={0.0cm 0.0cm 0.0cm 0.0cm}, clip=true, width=1.0\textwidth]{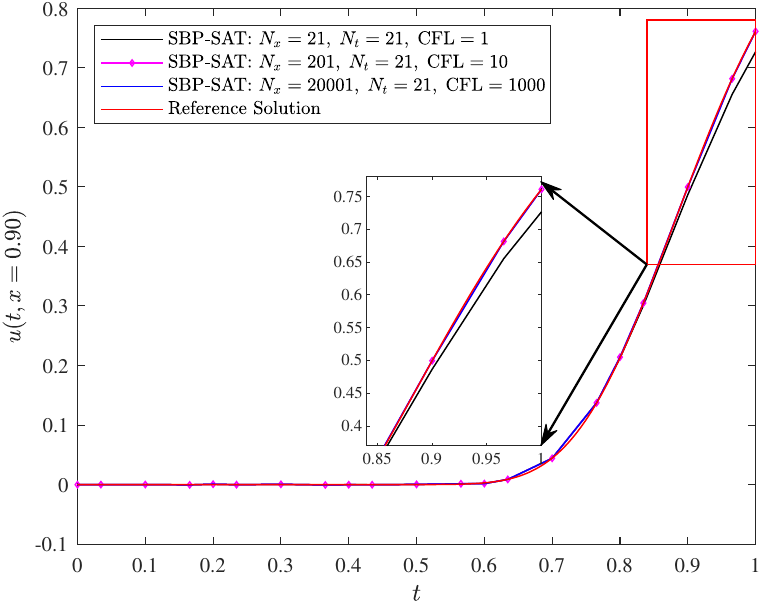}
\end{subfigure}%
 \begin{subfigure}[ht!]{0.5\textwidth}
		        \centering
		\includegraphics[trim={0.0cm 0.0cm 0.0cm 0.0cm}, clip=true, width=1.0\textwidth]{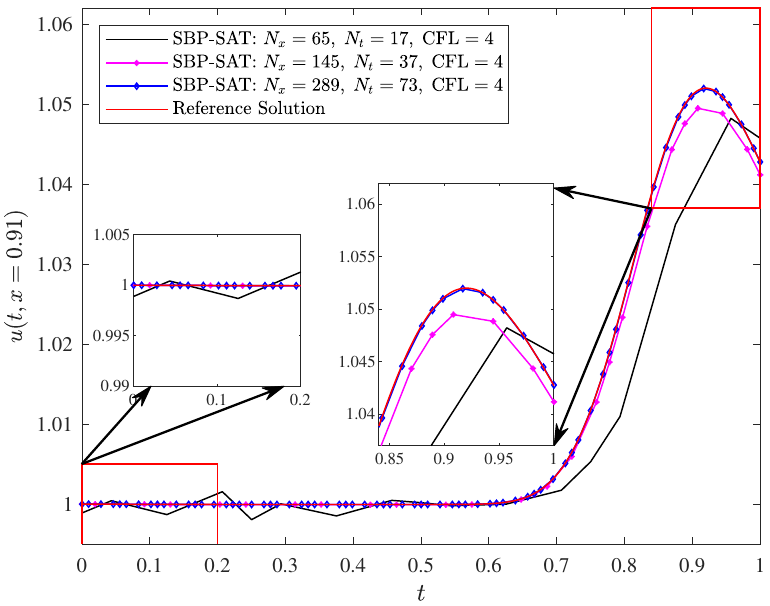}
\end{subfigure}
    \caption{Convergence of solution at $x=0.9$ for $g=1$ (left) and (right) $g(t)$ takes the form of (\ref{Eq:InflowBC_timeVarying_g}) }
\label{Fig:WavePropagation_SBP_spaceTime_atFix_x}
    \end{figure*}      
    
To analyze the efficiency of the combined space-time SBP method, we employ Lagrange polynomials of degree 4 in both spatial and temporal directions, which produce a $6^{\textrm{th}}$ order accurate numerical scheme in either direction. We first examine the mesh convergence behavior for the steady-state problem by comparing the results with the reference solution. The results are presented in the left part of Figure~\ref{Fig:WavePropagation_SBP_spaceTime}, which shows the $u$-profile at the final time $t=1$ as a function of $x$. Oscillations in the solution are observed near the boundary layer when coarse spatial meshes are used. These boundary layer oscillations vanish with the finer spatial meshes. The results show that the developed numerical scheme produces accurate results at very high CFL numbers. This highlights the efficiency of our proposed approach for steady state problems.

Next, we consider the time-dependent case. The corresponding mesh convergence studies are  shown in the right part of the Figure~\ref{Fig:WavePropagation_SBP_spaceTime}. It can be seen that the numerical solutions obtained using our fully discrete SBP method monotonically converges to the reference solution with the mesh refinement.

We next examine the temporal convergence of the solution at $x=0.9$, located within the boundary layer, for several mesh resolutions. The results are shown in Figure~\ref{Fig:WavePropagation_SBP_spaceTime_atFix_x}. For both the steady-state and time-dependent problems, the solutions obtained on relatively coarse meshes are in close agreement with the fine-mesh reference solutions. Moreover, as the mesh is refined, the numerical solutions exhibit a monotonic convergence toward the reference solutions.
\begin{table*}[ht!]
\centering
\caption{Detailed mesh properties for the cost-efficiency studies }
\label{tab:meshProperties_costEfficiencyComp}
\resizebox{0.52\textwidth}{!}{%
\begin{tabular}{cccccccc}
\hline
\multicolumn{8}{c}{Mesh properties}                                                                                                    \\ \hline
\multicolumn{5}{c}{SBP-SAT scheme}                                            & \multicolumn{3}{c}{RK4 explicit}                   \\ \hline
\multicolumn{2}{c}{Spatial} & \multicolumn{2}{c}{Temporal} & \multirow{2}{*}{CFL} & \multicolumn{2}{c}{Spatial} & \multirow{2}{*}{CFL} \\ \cmidrule(lr){1-4} \cmidrule(lr){6-7}
$N_x$            & $p_x$          & $N_t$            & $p_t$           &                      & $N_x$            & $p_x$          &                      \\ \hline
601           & 4           & 201           & 4            & 4.0683               & 601           & 4           & 0.030                \\
901           & 4           & 301           & 4            & 4.0683               & 901           & 4           & 0.023                \\
1201          & 4           & 401           & 4            & 4.0683               & 1201          & 4           & 0.016                \\
1501          & 4           & 501           & 4            & 4.0683               & 1501          & 4           & 0.010                \\ \hline
\end{tabular}%
}
\end{table*}

 To study the efficiency more quantitatively, we have conducted cost comparison studies with the traditional RK4 technique for three different frequencies $\omega=5\pi,10\pi$ and $20\pi$. For each case, we have employed similar spatial meshes, described in  Table~\ref{tab:meshProperties_costEfficiencyComp}. 
 \begin{figure*}[ht!]
		        \centering
		 \begin{subfigure}[h!]{0.32\textwidth}
        \centering
        \includegraphics[trim={0.0cm 0.0cm 0.0cm 0.0cm}, clip=true, width=1.0\textwidth]{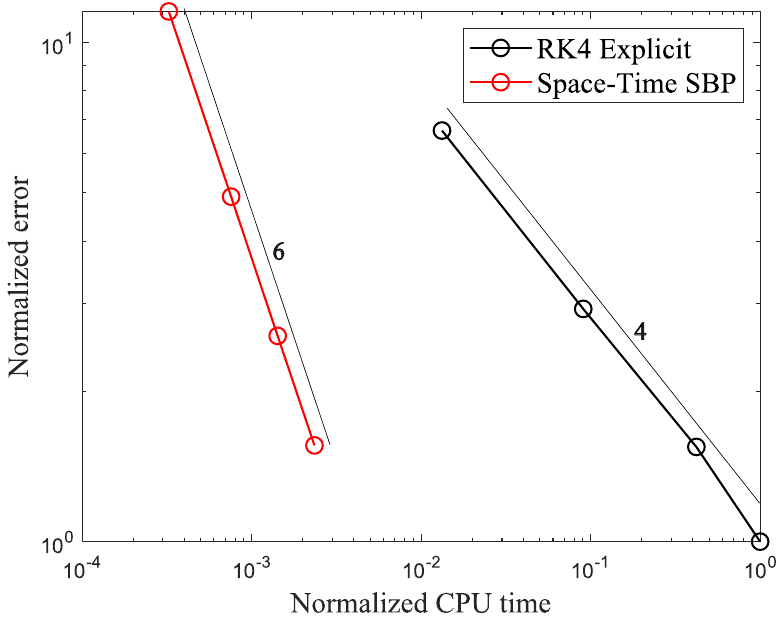}
\caption{}
\label{alpha_zer0}
\end{subfigure}%
 \begin{subfigure}[ht!]{0.32\textwidth}
		        \centering
		\includegraphics[trim={0.0cm 0.0cm 0.0cm 0.0cm}, clip=true, width=1.0\textwidth]{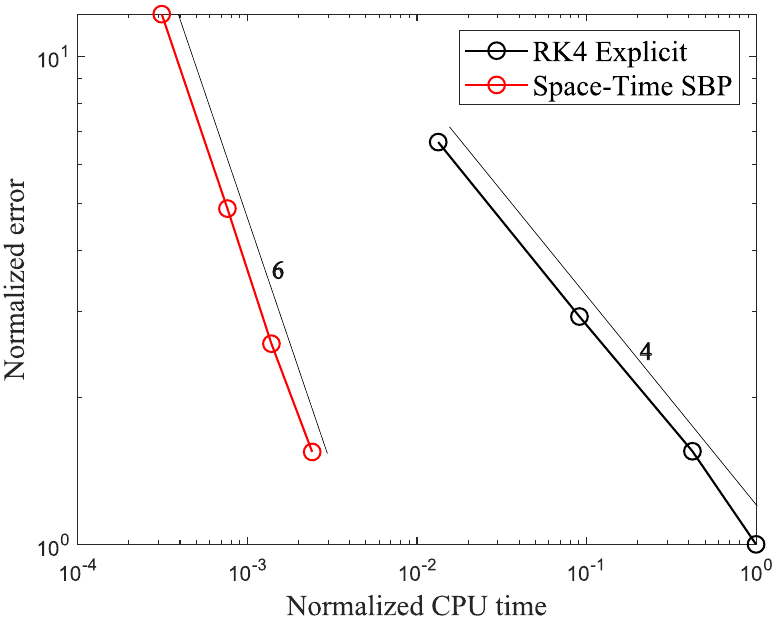}
\caption{}
\label{alpha_5em1}
\end{subfigure}%
 \begin{subfigure}[ht!]{0.32\textwidth}
		        \centering
		\includegraphics[trim={0.0cm 0.0cm 0.0cm 0.0cm}, clip=true, width=1.0\textwidth]{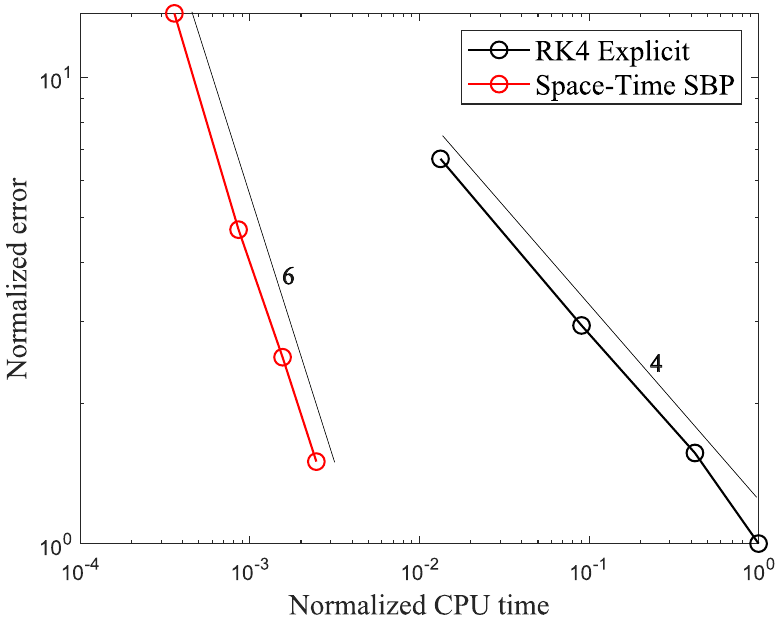}
\caption{}
\label{alpha_1}
\end{subfigure}
    \caption{Cost-efficiency studies for amplitude $\alpha=1$ with various frequencies (a) $\omega=5\pi$, (b) $\omega=10\pi$, and (c) $\omega=20\pi$   }
\label{Fig:Cost_efficiency_comp_VarFreq_SBP_RK4}
    \end{figure*}
The results are presented in Figure~\ref{Fig:Cost_efficiency_comp_VarFreq_SBP_RK4}. The error is computed relative to the reference solution described above. The figure illustrates the normalized error as a function of the normalized CPU time. Both the error and CPU time are normalized with respect to the RK4 results obtained using $N_x = 1501$ and a CFL number of 0.01. For the implicit space-time SBP method very coarse meshes are used in the temporal directions, whereas the RK4 method requires a very fine temporal mesh for stability. Figure~\ref{Fig:Cost_efficiency_comp_VarFreq_SBP_RK4} shows that the RK4 method carries very high computational cost compared to the SBP space-time method for all cases which highlights the computational efficiency of the implicit scheme. Roughly speaking the SBP-SAT scheme is between two to three orders of magnitude more efficient.
\section{Summary and Conclusions}
\label{sec:Concluding_remarks}
A high-order accurate fully discrete CG formulation in the SBP-SAT framework has been developed to solve initial-boundary value problems for advection-diffusion like equations. The energy analysis has been conducted at both the continuous, semi-discrete, and fully discrete levels, proving the stability of the method through the establishment of an energy bound.

Numerical validation by the method of manufactured solutions confirm superconvergence in both space and time with an order of accuracy $\mathcal{O}(p+2)$, for $p\geq 2$. In an application case, the fully discrete formulation efficiently captures space-time variations even on coarse meshes, demonstrating the method’s computational effectiveness. The application to nonlinear problems is kept for future studies.
\subsubsection*{Acknowledgements}
 M. Mandal and A. G. Malan were financially supported by the National Research Foundation (NRF) of South Africa grant 89916\footnote{Any opinion, findings and conclusions or recommendations expressed in this material are those of the author(s) and therefore the NRF and DST do not accept any liability with regard thereto.} as well as by the European Union HASTA -101138003. J. Nordström was funded by Vetenskapsrådet Sweden grant 2021-05484, University of Johannesburg Global Excellence and Stature Initiative Funding and the NRF of South Africa grant 89916.
 
\section*{Statements and Declarations}
\subsubsection*{Competiting Interests}
The authors declare no competing interests.

\subsubsection*{Data Availability}
The data used in this study are described in the article. Any additional data that may be required are available upon request.


\begin{thebibliography}{19}
\expandafter\ifx\csname natexlab\endcsname\relax\def\natexlab#1{#1}\fi
\providecommand{\url}[1]{\texttt{#1}}
\providecommand{\path}[1]{#1}
\providecommand{\DOIprefix}{doi:}
\providecommand{\ArXivprefix}{arXiv:}
\providecommand{\URLprefix}{URL: }
\providecommand{\Pubmedprefix}{pmid:}
\providecommand{\doi}[1]{\href{http://dx.doi.org/#1}{\path{#1}}}
\providecommand{\Pubmed}[1]{\href{pmid:#1}{\path{#1}}}
\providecommand{\bibinfo}[2]{#2}
\ifx\xfnm\relax \def\xfnm[#1]{\unskip,\space#1}\fi
\bibitem[{Sv{\"a}rd and Nordstr{\"o}m(2014)}]{svard2014review}
\bibinfo{author}{M.~Sv{\"a}rd}, \bibinfo{author}{J.~Nordstr{\"o}m},
\newblock \bibinfo{title}{Review of summation-by-parts schemes for
  initial--boundary-value problems},
\newblock \bibinfo{journal}{Journal of Computational Physics}
  \bibinfo{volume}{268} (\bibinfo{year}{2014}) \bibinfo{pages}{17--38}.
\bibitem[{Fern{\'a}ndez et~al.(2014)Fern{\'a}ndez, Hicken, and
  Zingg}]{fernandez2014review}
\bibinfo{author}{D.~C. D.~R. Fern{\'a}ndez}, \bibinfo{author}{J.~E. Hicken},
  \bibinfo{author}{D.~W. Zingg},
\newblock \bibinfo{title}{Review of summation-by-parts operators with
  simultaneous approximation terms for the numerical solution of partial
  differential equations},
\newblock \bibinfo{journal}{Computers \& Fluids} \bibinfo{volume}{95}
  (\bibinfo{year}{2014}) \bibinfo{pages}{171--196}.
\bibitem[{Hanif et~al.(2025)Hanif, Nordstr{\"o}m, Malan
  et~al.}]{hanif2025efficiency}
\bibinfo{author}{H.~Hanif}, \bibinfo{author}{J.~Nordstr{\"o}m},
  \bibinfo{author}{A.~Malan}, et~al.,
\newblock \bibinfo{title}{Efficiency analysis of continuous and discontinuous
  {G}alerkin finite element methods},
\newblock \bibinfo{journal}{AIMS Mathematics} \bibinfo{volume}{10}
  (\bibinfo{year}{2025}) \bibinfo{pages}{22579--22597}.
\bibitem[{Nordstr{\"o}m and Lundquist(2013)}]{nordstrom2013summation}
\bibinfo{author}{J.~Nordstr{\"o}m}, \bibinfo{author}{T.~Lundquist},
\newblock \bibinfo{title}{Summation-by-parts in time},
\newblock \bibinfo{journal}{Journal of Computational Physics}
  \bibinfo{volume}{251} (\bibinfo{year}{2013}) \bibinfo{pages}{487--499}.
\bibitem[{Lundquist and Nordstr{\"o}m(2014)}]{lundquist2014sbp}
\bibinfo{author}{T.~Lundquist}, \bibinfo{author}{J.~Nordstr{\"o}m},
\newblock \bibinfo{title}{The {SBP-SAT} technique for initial value problems},
\newblock \bibinfo{journal}{Journal of Computational Physics}
  \bibinfo{volume}{270} (\bibinfo{year}{2014}) \bibinfo{pages}{86--104}.
\bibitem[{Kennedy and Carpenter(2003)}]{kennedy2003additive}
\bibinfo{author}{C.~A. Kennedy}, \bibinfo{author}{M.~H. Carpenter},
\newblock \bibinfo{title}{Additive {R}unge--{K}utta schemes for
  convection--diffusion--reaction equations},
\newblock \bibinfo{journal}{Applied numerical mathematics} \bibinfo{volume}{44}
  (\bibinfo{year}{2003}) \bibinfo{pages}{139--181}.
\bibitem[{Carpenter et~al.(2005)Carpenter, Kennedy, Bijl, Viken, and
  Vatsa}]{carpenter2005fourth}
\bibinfo{author}{M.~H. Carpenter}, \bibinfo{author}{C.~A. Kennedy},
  \bibinfo{author}{H.~Bijl}, \bibinfo{author}{S.~Viken}, \bibinfo{author}{V.~N.
  Vatsa},
\newblock \bibinfo{title}{Fourth-order {R}unge-{K}utta schemes for fluid
  mechanics applications},
\newblock \bibinfo{journal}{Journal of Scientific Computing}
  \bibinfo{volume}{25} (\bibinfo{year}{2005}) \bibinfo{pages}{157--194}.
\bibitem[{Cash(1983)}]{cash1983integration}
\bibinfo{author}{J.~R. Cash},
\newblock \bibinfo{title}{The integration of stiff initial value problems in
  {ODE}s using modified extended backward differentiation formulae},
\newblock \bibinfo{journal}{Computers \& mathematics with applications}
  \bibinfo{volume}{9} (\bibinfo{year}{1983}) \bibinfo{pages}{645--657}.
\bibitem[{Cash(2000)}]{cash2000modified}
\bibinfo{author}{J.~Cash},
\newblock \bibinfo{title}{Modified extended backward differentiation formulae
  for the numerical solution of stiff initial value problems in {ODE}s and
  {DAE}s},
\newblock \bibinfo{journal}{Journal of Computational and Applied Mathematics}
  \bibinfo{volume}{125} (\bibinfo{year}{2000}) \bibinfo{pages}{117--130}.
\bibitem[{Hundsdorfer and Ruuth(2006)}]{hundsdorfer2006monotonicity}
\bibinfo{author}{W.~Hundsdorfer}, \bibinfo{author}{S.~Ruuth},
\newblock \bibinfo{title}{On monotonicity and boundedness properties of linear
  multistep methods},
\newblock \bibinfo{journal}{Mathematics of Computation} \bibinfo{volume}{75}
  (\bibinfo{year}{2006}) \bibinfo{pages}{655--672}.
\bibitem[{Hundsdorfer et~al.(2012)Hundsdorfer, Mozartova, and
  Spijker}]{hundsdorfer2012stepsize}
\bibinfo{author}{W.~Hundsdorfer}, \bibinfo{author}{A.~Mozartova},
  \bibinfo{author}{M.~Spijker},
\newblock \bibinfo{title}{Stepsize restrictions for boundedness and
  monotonicity of multistep methods},
\newblock \bibinfo{journal}{Journal of Scientific Computing}
  \bibinfo{volume}{50} (\bibinfo{year}{2012}) \bibinfo{pages}{265--286}.
\bibitem[{Costabile and Napoli(2001)}]{costabile2001method}
\bibinfo{author}{F.~Costabile}, \bibinfo{author}{A.~Napoli},
\newblock \bibinfo{title}{A method for global approximation of the initial
  value problem},
\newblock \bibinfo{journal}{Numerical Algorithms} \bibinfo{volume}{27}
  (\bibinfo{year}{2001}) \bibinfo{pages}{119--130}.
\bibitem[{Guo and Wang(2009)}]{guo2009legendre}
\bibinfo{author}{B.-y. Guo}, \bibinfo{author}{Z.-q. Wang},
\newblock \bibinfo{title}{Legendre--{G}auss collocation methods for ordinary
  differential equations},
\newblock \bibinfo{journal}{Advances in Computational Mathematics}
  \bibinfo{volume}{30} (\bibinfo{year}{2009}) \bibinfo{pages}{249--280}.
\bibitem[{Wang and Guo(2012)}]{wang2012legendre}
\bibinfo{author}{Z.-q. Wang}, \bibinfo{author}{B.-y. Guo},
\newblock \bibinfo{title}{Legendre-{G}auss-{R}adau collocation method for
  solving initial value problems of first order ordinary differential
  equations},
\newblock \bibinfo{journal}{Journal of Scientific Computing}
  \bibinfo{volume}{52} (\bibinfo{year}{2012}) \bibinfo{pages}{226--255}.
\bibitem[{Strand(1994)}]{strand1994summation}
\bibinfo{author}{B.~Strand},
\newblock \bibinfo{title}{Summation by parts for finite difference
  approximations for $d/dx$},
\newblock \bibinfo{journal}{Journal of Computational Physics}
  \bibinfo{volume}{110} (\bibinfo{year}{1994}) \bibinfo{pages}{47--67}.
\bibitem[{Nordstr{\"o}m(2017)}]{nordstrom2017roadmap}
\bibinfo{author}{J.~Nordstr{\"o}m},
\newblock \bibinfo{title}{A roadmap to well posed and stable problems in
  computational physics},
\newblock \bibinfo{journal}{Journal of Scientific Computing}
  \bibinfo{volume}{71} (\bibinfo{year}{2017}) \bibinfo{pages}{365--385}.
\bibitem[{Arnold et~al.(2002)Arnold, Brezzi, Cockburn, and
  Marini}]{arnold2002unified}
\bibinfo{author}{D.~N. Arnold}, \bibinfo{author}{F.~Brezzi},
  \bibinfo{author}{B.~Cockburn}, \bibinfo{author}{L.~D. Marini},
\newblock \bibinfo{title}{Unified analysis of discontinuous {G}alerkin methods
  for elliptic problems},
\newblock \bibinfo{journal}{SIAM journal on numerical analysis}
  \bibinfo{volume}{39} (\bibinfo{year}{2002}) \bibinfo{pages}{1749--1779}.
\bibitem[{Malan and Nordstrom(2023)}]{malan2023sbp}
\bibinfo{author}{A.~G. Malan}, \bibinfo{author}{J.~Nordstrom},
\newblock \bibinfo{title}{An {SBP-SAT} continuous {G}alerkin {F}inite {E}lement
  formulation for smooth and discontinuous fields},
\newblock \bibinfo{journal}{arXiv preprint arXiv:2311.05395}
  (\bibinfo{year}{2023}).
\bibitem[{Roache(2002)}]{roache2002code}
\bibinfo{author}{P.~J. Roache},
\newblock \bibinfo{title}{Code verification by the method of manufactured
  solutions},
\newblock \bibinfo{journal}{J. Fluids Eng.} \bibinfo{volume}{124}
  (\bibinfo{year}{2002}) \bibinfo{pages}{4--10}.

\end{thebibliography}
\end{document}